\documentclass[%
 preprint,
superscriptaddress,
 amsmath,amssymb,
 aps,
prb,
]{revtex4-2}

\usepackage[english]{babel}
\usepackage[utf8]{inputenc}
\usepackage{amsthm}
\usepackage{mathtools}
\usepackage{physics}
\usepackage{xcolor}
\usepackage{graphicx}
\usepackage[left=23mm,right=13mm,top=35mm,columnsep=15pt]{geometry} 
\usepackage{adjustbox}
\usepackage{placeins}
\usepackage[T1]{fontenc}
\usepackage{lipsum}
\usepackage{csquotes}\usepackage[pdftex, pdftitle={Article}, pdfauthor={Author}]{hyperref} 
\usepackage{isotope}
\begin{document}
\title{Local magnetic and geometric structure in Mn-doped La(Fe,Si)$_{\text{13}}$}

\author{Benedikt Eggert}
    \email[Correspondence email address: ]{Benedikt.Eggert@uni-due.de}
    \affiliation{Faculty of Physics and Center for Nanointegration Duisburg-Essen (CENIDE), University of Duisburg-Essen, Lotharstr.1, 47057 Duisburg, Germany}
\author{Johanna Lill}
    \affiliation{Faculty of Physics and Center for Nanointegration Duisburg-Essen (CENIDE), University of Duisburg-Essen, Lotharstr.1, 47057 Duisburg, Germany}
\author{Damian Günzing}
    \affiliation{Faculty of Physics and Center for Nanointegration Duisburg-Essen (CENIDE), University of Duisburg-Essen, Lotharstr.1, 47057 Duisburg, Germany}
\author{Cynthia Pillich}
    \affiliation{Faculty of Physics and Center for Nanointegration Duisburg-Essen (CENIDE), University of Duisburg-Essen, Lotharstr.1, 47057 Duisburg, Germany}
\author{Alexandra Terwey}
    \affiliation{Faculty of Physics and Center for Nanointegration Duisburg-Essen (CENIDE), University of Duisburg-Essen, Lotharstr.1, 47057 Duisburg, Germany}
\author{Ilyia A. Radulov}
    \affiliation{Functional Materials, Institute of Materials Science, Technische Universität, 64287 Darmstadt, Germany}
\author{Fabrice Wilhelm}
    \affiliation{European Synchrotron Radiation Facility, 38043 Grenoble cedex, France}
\author{Andrei Rogalev}
    \affiliation{European Synchrotron Radiation Facility, 38043 Grenoble cedex, France}
\author{Mauro Rovezzi}
    \affiliation{Univ. Grenoble Alpes, CNRS, IRD, Irstea, Météo France, OSUG, FAME, 38000 Grenoble, France}
\author{Konstantin Skokov}
    \affiliation{Functional Materials, Institute of Materials Science, Technische Universität, 64287 Darmstadt, Germany}
\author{Katharina Ollefs}
    \affiliation{Faculty of Physics and Center for Nanointegration Duisburg-Essen (CENIDE), University of Duisburg-Essen, Lotharstr.1, 47057 Duisburg, Germany}
\author{Oliver Gutfleisch}
    \affiliation{Functional Materials, Institute of Materials Science, Technische Universität, 64287 Darmstadt, Germany}
\author{Markus E. Gruner}
    \affiliation{Faculty of Physics and Center for Nanointegration Duisburg-Essen (CENIDE), University of Duisburg-Essen, Lotharstr.1, 47057 Duisburg, Germany}

\author{Heiko Wende}
    \affiliation{Faculty of Physics and Center for Nanointegration Duisburg-Essen (CENIDE), University of Duisburg-Essen, Lotharstr.1, 47057 Duisburg, Germany}

\date{\today} 

\begin{abstract}
Magnetic cooling has the potential to replace conventional gas compression refrigeration. Materials such as La(Fe,Si)$_{13}$ exhibit a sizeable first-order magnetocaloric effect, and it is possible to tailor the phase transition towards room temperature by  Mn-H-doping, resulting in a large temperature range for operation. Within this work, we discuss variations of the electronic and lattice structure in La(Fe,Si)$_{13}$ with increasing Mn content utilizing X-ray magnetic circular dichroism (XMCD) and extended X-ray absorption fine structure spectroscopy (EXAFS). While XMCD shows a decrease of the magnetic polarization at the Fe K edge, low-temperature EXAFS measurements indicate increased positional disorder in the La environment that is otherwise absent for Fe and Mn. First-principles calculations link the positional disorder to an enlarged Mn-Si distance -- explaining the increased positional disorder in the La surrounding.
\end{abstract}

\maketitle

\section{Introduction}

		The discovery of the giant magnetocaloric (MC) effect by Pe\-char\-sky and Gschneider in Gd$_5$Ge$_2$Si$_2$ in 1997~\cite{Pecharsky1997} started the research field of magnetic refrigeration, which aims for an en\-vi\-ron\-men\-tal\-ly friend\-ly replacement of gas-compression refrigeration cycles~\cite{Benke2020,Benke2021}. The materials of interest exhibit a significant adiabatic temperature change $\Delta T_\text{ad}$ induced by external applied magnetic fields and isothermal entropy change $\Delta S_\text{iso}$ at their first-order phase transition~\cite{Gutfleisch2016}. 
	
	Besides the benchmark systems FeRh~\cite{Barua2014,SternTaulats2017,Tran2022} and Gd\cite{Pecharsky1997,Dankov1998,Bahl2009}, current materials of interest~\cite{Sandeman2012,Scheibel2018,Gottschall2019,Zarkevich2020}, possessing a sizeable magnetocaloric effect near room temperature, are, e.g. NiMn-based Heusler compounds~\cite{Krenke2005,Maosa2010,PrezLandazbal2017}, MnFePSi~\cite{Tegus2002,CamThanh2008,Dung2011,Guillou2014,Funk2018}, and H-containing La(Fe,Si)$_{13}$~\cite{Fukamichi2006,Fujita2009,Shen2009,Liu2011,Fujita2012,Maosa2011,Imaizumi2022}. Especially the latter system seems suitable for room temperature applications due to its small thermal hysteresis width and the possibility of tailoring the phase transition by adjusting the Si and H concentration. To further reduce the thermal hysteresis and increase the cooling efficiency towards the theoretically predicted ones, developing new mechanisms to tailor the hysteresis of the first-order phase transition of these magnetocaloric compounds is essential. To implement these new mechanisms, a complete microscopic understanding of the electronic and magnetic interactions is of help~\cite{Morrison2022}.  
	
	The promising magnetocaloric system La(Fe,Si)$_{13}$H possesses a significant adiabatic temperature change $\Delta T_{\text{ad}}$ and a drastic volume decrease $\Delta V/V$~\cite{Jia2006} along its isostructural first-order phase transition, which can be correlated to an itinerant electron metamagnetic transition (IEM)~\cite{Fujita1999,Fujita2001}. The magnetocaloric compound crystallizes in the NaZn13 (Fm3c) structure (1:13 phase) with cubic shape. Here, two nonequivalent Fe sites are present, which we will refer to as Fe$_{\mathrm{I}}$ site (8b Wyckoff position) and Fe$_{\mathrm{II}}$ site (96i Wyckoff position). The Si atoms are assumed to occupy mainly the Fe$_{\mathrm{II}}$ sites, sharing this site with Fe atoms~\cite{Wang2003,Hamdeh2004,Rosca2010}. The unit cell can be represented by a cell with fcc primitive vectors containing 28 atoms~\cite{Jia2006,Kuzmin2007,Gercsi2017}, which simplifies computational studies. The precise ratio of Si and H in the La(Fe,Si)$_{13}$ system will determine the phase transition temperature, which is aimed to be around room temperature. A high Si content (above x = 1.8~\cite{Gutfleisch2005}) will turn the phase transition to second order, which reduces the magnetocaloric effect but avoids dissipative hysteresis losses. In the literature, many ongoing discussions debate the influence of different dopings~\cite{Jia2011} on the size of the magnetocaloric effect and, e.g. phase transition order or thermal hysteresis width. Examples are Cr-doping, which first decreases thermal hysteresis width by increasing Cr content to a concentration of x=0.3. In contrast, for higher doping, the thermal hysteresis increases again~\cite{MorenoRamrez2019}. Besides, Ni-doping changes the first-order phase transition character to second-order while leading to a rise of the phase transition temperature T$_\text{tr}$~\cite{MorenoRamrez2018}. While Si is necessary to stabilize the La(Fe,Si)$_{13}$ compound, H-doping shifts the phase transition by about 100\,K to RT~\cite{Fujita2003,Wang2003,Barcza2011,Krautz2014}. A combined nuclear resonant inelastic X-ray scattering and density functional theory investigation of the hydrogenated compound could show that the nature of the mechanism is identical to the parent compound~\cite{Gruner2015,Terwey2020}. Partial H-loading makes it possible to tune the phase transition towards 300\,K, while having the disadvantage that partial H-doping leads to phase separation into a hydrogen-rich and a hydrogen-poor phase, leading towards the distribution of phase transition temperatures~\cite{Baumfeld2014}. A complete hydrogen loading can overcome this, but the phase transition occurs at $T=360\,\mathrm{K}$ for these high H-contents. This makes it necessary to dope the system further in order to have a system with a phase transition temperature at 300\,K and overcoming the potential risk of phase separation by the introduction of Mn in the NaZn$_{13}$structure~\cite{Krautz2014,Zhou2021}.
	
	The current work intends to resolve the occurring variations in interactions between the different lattice sites and elements to gain a microscopic picture of the Mn-doped compound. First, we investigate the La and Fe's magnetic moments in the ferromagnetic ground state with increasing Mn concentration by employing X-ray magnetic circular dichroism (XMCD) in the hard X-ray regime and compare these results with conventional magnetometry measurements. In a second step, we discuss changes in the local structure employing extended X-Ray absorption fine structure spectroscopy (EXAFS) at the Fe, Mn and La edges, enabling us to compare the local environment and how this is affected by Mn doping for the different lattice sites. These experimental investigations are complemented by \textit{ab-initio} calculations to attribute structural changes to selected bonds and reveal changes off the respective magnetic moment. Based on this approach, we can show that by introducing Mn in the NaZn$_{13}$-structure, a decrease of the magnetisation and the magnetic moment of Fe occurs, combined with an increased structural disorder within the first backscattering shell around the La-site. First-principles calculations and EXAFS measurements indicate that Mn preferentially occupies the 96i sites, while first-principles calculations indicate the presence of different local environments of Mn.

	\begin{figure*}[h!]
		\centering
		\includegraphics[width=\textwidth]{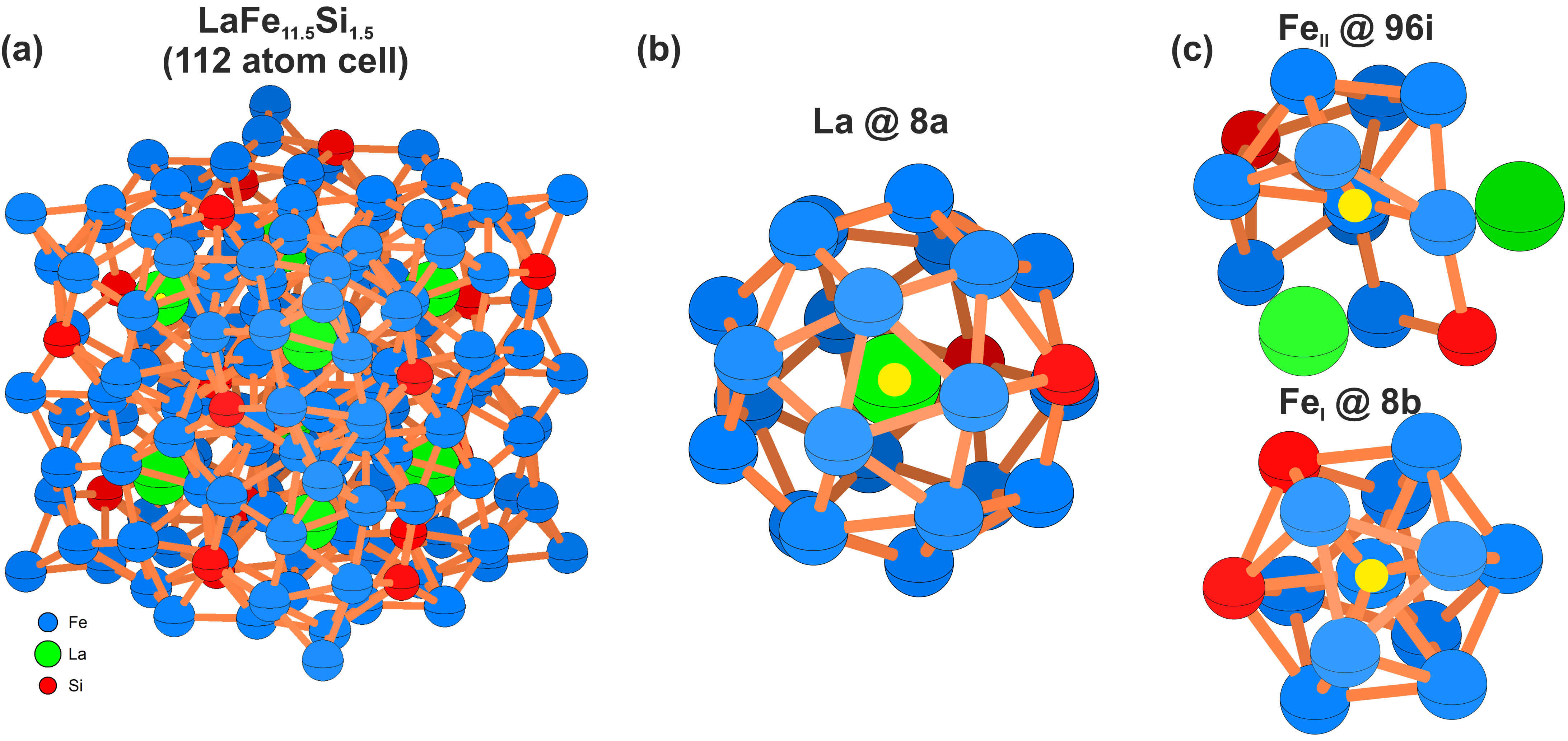}
		\caption{(a) 112 atom cell of LaFe$_{\text{11.5}}$Si$_{\text{1.5}}$ in the NaZn$_{\text{13}}$ structure. Exemplary representation along the (111)-direction of (b) La and (c) Fe surrounding occupying the 8a, 8b (Fe$_{\text{I}}$), and 96i (Fe$_{\text{II}}$)-site. Surroundings depict the local environment within a distance of 3.3\,\AA~(La), 3.3\,\AA~(Fe@Fe$_{\text{I}}$), and 4.0\,\AA~(Fe@Fe$_{\text{II}}$), respectively. The yellow circle illustrates the central absorbing atom used in the respective calculations.} 
		\label{fig:LaFeSi-Structure}
	\end{figure*}
	\section{Experimental details}
	
	Mn-containing LaFe$_{\text{13-x-y}}$Mn$_{\text{x}}$Si$_{\text{y}}$ with a fixed Si content of y=1.4 and a Mn content ranging from x=0 to x=0.5 were prepared by induction melting in an argon atmosphere from pure elements. Afterwards, the as-cast samples were homogenised by annealing at 1323\,K for seven days in an Ar-filled and sealed quartz ampoule, followed by quenching in water. The obtained ingots were then crushed into a powder for further measurements. We refer the reader to Refs.~\cite{Liu2011,Krautz2014} for additional details on the sample preparation. Field-dependent magnetometry was performed using a Quantum Design PPMS DynaCool with the vibrating sample magnetometer option applying an external magnetic field up to 9\,T with a sample temperature range between 1.8\,K up to 400\,K. Element-specific extended X-ray absorption fine structure (EXAFS) measurements were conducted at beamline BM30B at the ESRF~\cite{BM30B}. Spectra have been recorded at the corresponding K or L$_3$ absorption edge in partial fluorescence yield detection mode at 45° incident geometry using an energy dispersive detector to collect the signal originating from the respective characteristic K$_{\alpha}$ or L$_{\alpha}$ fluorescence line. In addition, potential saturation effects due to self-absorption have been corrected using the Booth algorithm~\cite{Booth2005} within the DEMETER package~\cite{Ravel2005}. X-ray magnetic circular dichroism (XMCD) measurements were conducted at beamline ID12 at the ESRF at low temperatures (T=3\,K) with an external magnetic field of $\pm$2\,T applied parallel to the incoming X-ray beam. The obtained XMCD spectra arise as the difference between X-ray absorption spectra measured with opposite helicities of the incoming photons. The measurements are performed in opposite directions of the magnetic field to avoid potential artefacts in the dichroic signal.    
	
	\section{Computational details}
	Parameter-free first-principles calculations were carried out in the framework of the density functional theory (DFT) with the help of the Vienna Ab-initio Simulation Package (VASP)~\cite{cn:VASP1}. A plane wave basis set with an energy-cutoff $E_\mathrm{ cut}=400\,$eV was used with exchange and correlation in the generalised gradient approximation with the functional of Perdew, Burke and Ernzerhof~\cite{cn:Perdew96} and scalar-relativistic approximation with collinear spin-polarisation. For the accurate description of the core electrons, we took advantage of the projector augmented wave (PAW) approach~\cite{cn:VASP2}, employing potentials to a valence electron configuration of 5s$^2$5p$^6$5d$^1$6s$^2$ for La, 5p$^6$3d$^7$4s$^1$ for Fe, 5p$^6$3d$^6$4s$^1$ for Mn and 3s$^2$3p$^2$ for Si. For the structural optimisations (cell parameters, angles and atomic positions) in the 28-atom cell as used in our previous work \cite{Gruner2015} were carried out with of 7$\times$7$\times$7 Monkhorst-Pack $k$-mesh which yields 172 $k$-points in the irreducible Brillouin zone (IBZ) and a finite temperature smearing according to Methfessel and Paxton~\cite{cn:Methfessel89} with a broadening of $\sigma=\,0.1\mathrm{\,eV}$. The self-consistency cycle was stopped when the difference in energy between two consecutive cycles fell below $10^{-7}$\,eV, structures were considered relaxed below a threshold of $10^{-7}\mathrm{\,eV}$. Total energy and magnetic moments were determined in a further step with a finer 11$\times$11$\times$11 k-mesh (666 points in the IBZ), and the tetrahedron method with Bloechl corrections~\cite{cn:Bloechl94} for the Brillouin zone integration.
	
	The program package FEFF9~\cite{Rehr2000,Rehr2009,Rehr2010} was used for the calculation of the oscillatory fine structure of the X-ray absorption spectra in the ferromagnetic phase ($T=$20\,K) at the La L$_3$, Mn K and Fe K edges. FEFF calculations have been performed on the 112 atom cell representation of LaFe$_{\text{11.5}}$Si$_{\text{1.5}}$ (see Fig.~\ref{fig:LaFeSi-Structure}) in the NaZn$_\text{13}$ lattice structure, by averaging over all different occurring lattice sites. To model the fine structure of Mn, we assumed the absorber to be Mn, while the other atoms had not been adjusted. Due to the similar backscattering amplitude and phase shift of Mn and Fe, this assumption is justified. For the description of occurring dynamic disorder (e.g. phonons \& zero-point vibrations), we used a Debye model with a Debye temperature of 380\,K~\cite{Landers2018} and a measurement temperature of 20\,K. Besides, an additional static mean square relative displacement $\sigma^2_{\text{stat}}$ of 0.001$\,\text{\AA}^2$ has been assumed to model the presence of static disorder (grain boundaries or lattice distortions). Additional descriptions of the first scattering shell of the La, Mn and Fe atoms have been performed with the ARTEMIS package~\cite{Ravel2005} and Larch~\cite{Newville2001,Newville2013}.

	\section{Results \& Discussion}
	
	\subsection{The magnetic structure}

\begin{figure}[h!]
		\centering
		\includegraphics[width=.75\linewidth]{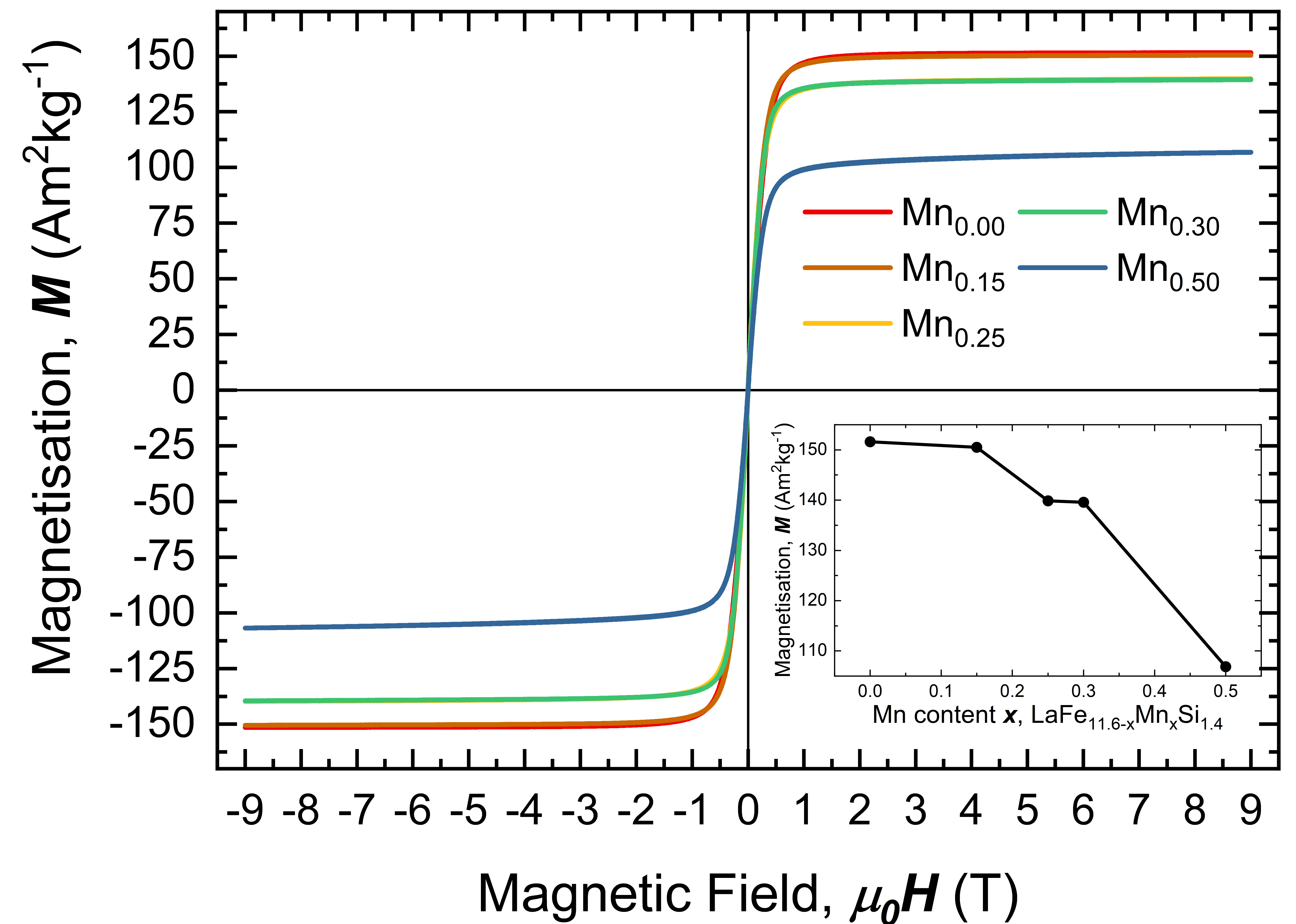}
		\caption{Field-dependent magnetometry performed at T = 5\,K for La(Fe,Si)$_{13}$ with varying Mn content. The obtained magnetisation values have been corrected with respect to occurring secondary bcc-Fe phases, as discussed in Ref.~\cite{Makarov2015}. The inset shows the obtained saturation magnetisation as a function of Mn concentration.}
		\label{fig:Magnetometrie}
	\end{figure}

	In order to track the effect of Mn-doping on the magnetic structure, we performed field-dependent magnetometry measurements. Figure~\ref{fig:Magnetometrie} depicts isothermal magnetisation measurements for different Mn-concentration between $\pm$9\,T at low temperatures. As shown in the inset of Figure~\ref{fig:Magnetometrie}, the saturation magnetisation reduces by approximately 10\,\% when Mn replaces 2.5\,\% of the Fe atoms while a Mn-concentration of approximately 4.5\,\% leads to a reduction of almost 30\,\% of the saturation magnetisation.
	
\begin{figure*}[h!]
		\centering
		\includegraphics[width=\linewidth]{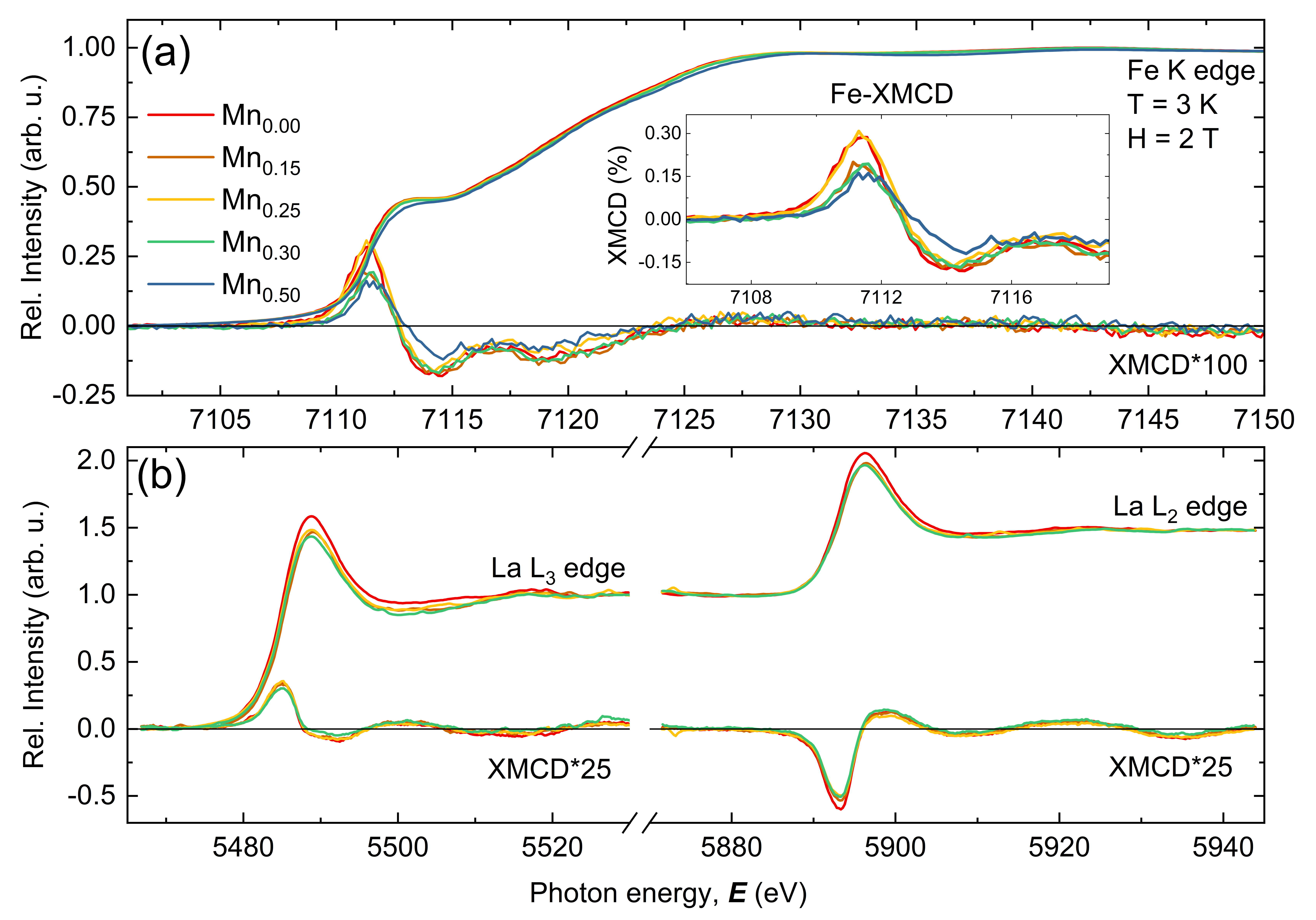}
		\caption{(a) X-ray absorption near edge spectra and X-ray magnetic circular dichroism signals for all LaFe$_{\text{11.6-x}}$Mn$_{\text{x}}$Si$_{\text{1.4}}$ samples at the Fe K edge, measured at 3\,K in an external field of 2\,T. Inset shows the near edge region of the Fe K edge XMCD signal. (b) X-ray absorption near edge spectra and X-ray magnetic circular dichroism signals for all except the maximum doped LaFe$_{\text{11.6-x}}$Mn$_{\text{x}}$Si$_{\text{1.4}}$ at the La L$_{2,3}$ edges.} 
		\label{fig:XMCD}
	\end{figure*}

	To resolve and attribute the changes in the magnetisation to the specific elements, we performed element-specific X-ray magnetic circular dichroism (XMCD) measurements in the hard X-ray regime at the beamline ID12 of the ESRF. For the Fe magnetic moment, we performed measurements at the Fe K edge in order to probe the 1s$\xrightarrow{}$4p transition (E1 transition), making it possible to obtain the magnetic dichroism arising from the orbital moment of the 4p states and the magnetic moment of the 3d-states through the quadrupolar transition (E2-transition). 
	Figure~\ref{fig:XMCD}(a) depicts the obtained Fe K edge XMCD spectra for different Mn concentrations. Focusing on the XMCD spectra (inset), we can see that with increasing Mn content, a reduction of the XMCD intensity occurs first at photons energy of 7111\,eV -- corresponding to the quadrupolar transitions into localised 3d states. Only the largest Mn concentration possesses an intensity reduction at higher photon energies (above 7112\,eV), indicating a decreased spin polarisation of delocalised 4p states. In addition, the XAS at photon energies of 7113\,eV shows a broadening close to the Fermi energy. The reduced Fe moment might originate from a redistribution in the density of states due to the hybridisation of Fe-Mn 3d states~\cite{Gercsi2015} and can also be assigned as a feature of the reduced magnetisation~\cite{Gruner2017}. 
	
	Besides, we performed measurements to unravel the magnetic dichroism of La 5d states by probing the 2p$\xrightarrow{}$5d transition at the La L$_{2,3}$ edges (see Figure~\ref{fig:XMCD}(b)). Usually, it is possible to extract the orbital and spin moment by the so-called sum rule analysis~\cite{Thole1992,Carra1993}. A quantitative discussion of the La 5d moments goes beyond the scope of this work, and the reasoning will be explained in the following. The La spectrum depicts a well-developed dichroic oscillating spectral fine structure -- leading to a significant error in the determination of the integrated XMCD signal. In addition, the quadrupolar transitions contributing to the isotropic absorption result in underestimating the extracted 5d moment. The obtained XMCD spectra reveal two aspects. First, the La 5d moments align antiparallel to the applied field (and the Fe-moment). At the same time, it seems that the magnetic dichroism and the isotropic absorption (number of unoccupied states) do not react to the Mn doping in the range of the investigated doping concentration. Based on the spectra, we can assume that the induced La moment is constant within experimental error. Concentrating on the XMCD shape, we observe that the XMCD peaks lay within the rising edge of the absorption edge -- indicating a derivative-like signal. The majority of the XMCD signal in the near edge region can be modelled (not shown here) by applying an energy shift of 50 and 120\,meV to the L$_3$ and L$_2$ edge, respectively. This energy shift within the La-partial density of states is caused by the 3d-5d hybridization leading to a slight exchange splitting in the spin-resolved density of states~\cite{Gruner2015}. Differences in the determined energy shifts reflect the varying lifetimes of the excited states at the respective absorption edges~\cite{Krause1979}. 
	
	Based on the results obtained by magnetometry, the Fe K, and La L$_{2,3}$ XMCD spectra, the decrease in the magnetisation can be attributed to a reduction of the Fe magnetic moment.
	
	\subsection{First-principles calculations}\label{ch:DFT}
	
	\begin{table*}[htb]
		\caption{
			Total energy $\Delta E_{\mathrm{tot}}$, total moment $\mu_{\mathrm{tot}}$ per formula unit (f.u.) and averaged element-resolved magnetic moments $\mu_i$ of LaFe$_{11.5}$Si$_{1.5}$ (first row) and LaFe$_{11}$Si$_{1.5}$Mn$_{0.5}$ given per atom as obtained from our DFT calculations. $n_{\mathrm{Si}}$ specifies the number of Si atoms in the closest neighbourhoods of Mn (below $3\,{\mathrm{\AA}}$) in the specific configuration. The total energy $\Delta E_{\mathrm{tot}}$ is specified per 28-atom primitive cell relative to the average total energy across all inequivalent configurations with Mn placed on the 96i sites. The first column specifies the sites at which the Mn is placed. This refers either to a specific position indicated by the letter in parentheses or the average (avg.).
		}
		\label{tab:DFT-moments}
		\centering
		\begin{tabular}{|l|l|r|r|r|rrrr|}
			\hline
			& & $\Delta E_{\mathrm{tot}}$ & $n_{\mathrm{Si}}$ & $\mu_{\mathrm{tot}}$ &  $\mu_{\mathrm{La}}$ &  $\mu_{\mathrm{Fe}}$&  $\mu_{\mathrm{Si}}$&  $\mu_{\mathrm{Mn}}$\\
			& Mn@ & (meV)  &  &($\mu_{\mathrm{B}}$/f.u.) &  \multicolumn{4}{c|}{($\mu_{\mathrm{B}}$/atom)}  \\ \hline
			LaFe$_{11.5}$Si$_{1.5}$ & - & - & - & 24.47 & -0.200 & 2.219 & -0.101 & -\\ \hline
			LaFe$_{11}$Si$_{1.5}$Mn$_{0.5}$ & 96i avg. & 0.0 & 1.3 & 21.31 & -0.181 & 2.142 & -0.087 & -2.598 \\[0.5ex]
			&96i(a) & -81.2 & 0 &  21.20 & -0.181 & 2.131 & -0.092 & -2.540 \\
			&96i(b) & -29.1 & 1 &20.97 & -0.183 & 2.125 & -0.087 & -2.851 \\
			&96i(c) &  17.7 & 2 & 21.36 & -0.177 & 2.146 & -0.088 & -2.587 \\
			&96i(d) & -81.0 & 2 &21.28 & -0.185 & 2.139 & -0.089 & -2.549 \\
			&96i(e) &  43.7 & 1 &21.36 & -0.179 & 2.143 & -0.087 & -2.534 \\
			&96i(f) & 112.8 & 1& 21.56 & -0.179 & 2.158 & -0.085 & -2.483 \\
			&96i(g) & 17.0 & 2& 21.41 & -0.185 & 2.153 & -0.080 & -2.641 \\\hline
			LaFe$_{11}$Si$_{1.5}$Mn$_{0.5}$ &8b avg. & 128.6 & 1.5 & 23.53 & -0.186 & 2.196 & -0.093 & 0.658 \\[0.5ex]
			&8b(a) & 114.4 & 3 & 23.39 & -0.190 & 2.197 & -0.091 & 0.420 \\
			&8b(b) & 142.9 & 0 & 23.67 & -0.181 & 2.195 & -0.094 & 0.895 \\ \hline
		\end{tabular}
	\end{table*}
	\begin{table*}[htb]
		\caption{
			Structural parameters of the LaFe$_{11.5}$Si$_{1.5}$ (first row) and LaFe$_{11}$Si$_{1.5}$Mn$_{0.5}$ (all other rows) configurations obtained from our DFT calculations, in terms of the (Average) volume per atom $V_{\mathrm{at}}$ and the averaged interatomic distances $d$ within the closest neighbour shells (below $4\,$\AA{} for La and $3\,$\AA{} for Fe and Mn) between specified pairs of elements and all atoms (all).
		}
		\label{tab:DFT-dist}
		\centering
		\begin{tabular}{|l|r|rrrr|rrrr|}
			\hline
			& $V_{\mathrm{at}}$ & $d$(Mn-Si) & $d$(Fe$_{\mathrm{II}}$-Si) & $d$(Fe$_{\mathrm{I}}$-Si)  & $d$(La-Si) & $d$(La-all) & $d$(Fe$_{\mathrm{I}}$-all) & $d$(Fe$_{\mathrm{II}}$-all) & $d$(Mn-all)\\
			Mn@ & (\AA{}$^3$/at.)  &  \multicolumn{4}{c|}{(\AA)}  &  \multicolumn{4}{c|}{(\AA)}  \\ \hline
			- & 13.281 & - & 2.4948 & 2.4034 & 3.3781 & 3.3328 & 2.4505 & 2.5099 & - \\ \hline
			96i avg. & 13.236 & 2.5566 & 2.4952 & 2.4086 & 3.3691 & 3.3296 & 2.4459 & 2.5071 & 2.5082\\[0.5ex]
			96i(a) & 13.214 & -      & 2.4919 & 2.4043 & 3.3713 & 3.3281 & 2.4454 & 2.5071 & 2.4848\\
			96i(b) & 13.226 & 2.6015 & 2.4926 & 2.3958 & 3.3719 & 3.3283 & 2.4462 & 2.5065 & 2.5280 \\
			96i(c) & 13.244 & 2.5626 & 2.4901 & 2.4218 & 3.3752 & 3.3296 & 2.4482 & 2.5076 & 2.4928 \\
			96i(d) & 13.225 & 2.6261 & 2.4866 & 2.4229 & 3.3504 & 3.3279 & 2.4454 & 2.5052 & 2.5124\\
			96i(e) & 13.228 & 2.4998 & 2.5103 & 2.4058 & 3.3587 & 3.3282 & 2.4444 & 2.5074 & 2.4988 \\
			96i(f) & 13.262 & 2.5375 & 2.4990 & 2.4114 & 3.3672 & 3.3332 & 2.4456 & 2.5075 & 2.5082 \\
			96i(g) & 13.256 & 2.5119 & 2.4956 & 2.3983 & 3.3889 & 3.3321 & 2.4462 & 2.5087 & 2.5327 \\\hline
			8b avg. & 13.247 & 2.4210 & 2.4940 & 2.4083 & 3.3683 & 3.3308 & 2.4505 & 2.5066 & 2.4422 \\[0.5ex]
			8b(a) & 13.243 & 2.4210 & 2.4921 & -  &  3.3657 & 3.3302 & 2.4653 & 2.5064 & 2.4270 \\
			8b(b) & 13.250 & -      & 2.4958 & 2.4083 & 3.3708 & 3.3313 & 2.4357 & 2.5068 & 2.4573\\ \hline
		\end{tabular}
	\end{table*}
	
	To obtain a microscopic understanding of the site-preference of Mn and its impact on the magnetic and structural properties, we carried out DFT calculations of LaFe$_{11}$Si$_{1.5}$Mn$_{0.5}$ in its 28-atom primitive cell containing two formula units. The primitive cell is spanned by face-centered cubic lattice vectors and represents $1/8$ of the conventional 112 atom cell shown in Fig.\ \ref{fig:LaFeSi-Structure}. The Si atoms were distributed across the 96i in an arrangement which preserves rhombohedral long-range symmetry. This distribution allowed us previously to provide computationally efficient modelling of the vibrational density of states of ferro- and paramagnetic LaFe$_{11.5}$Si$_{1.5}$ yielding an excellent agreement with experiment \cite{Gruner2015,Gruner2017,Landers2018,Terwey2020}. The remaining rhombohedral symmetry in the cell leaves 7 inequivalent three-fold degenerate Fe$_{\mathrm{II}}$ on the 96i positions and two inequivalent Fe$_{\mathrm{I}}$ sites on the 8b positions of the 28-atom prototype cell with cubic symmetry. In the course of our calculations, we systematically replaced one Fe-atom on these nine inequivalent sites with one Mn and evaluated the changes in magnetic properties and lattice structure.

	Comparing the total energies, we find a clear preference for Mn to be located on the 96i sites, see Table \ref{tab:DFT-moments}, the average across all 7 inequivalent choices being almost 130\,meV below the average over Mn placed on the 8b sites, which is a considerable difference. Thus, while it cannot be excluded that during heat treatment at elevated temperatures some Mn atoms will replace Fe$_{\mathrm{I}}$, the vast majority of the Mn atoms can be expected to reside on the positions of Fe$_{\mathrm{II}}$. Among the Mn placed on the 96i sites, there is a large spread regarding the total energy of the respective configurations. While for two configurations, 96i(a) and 96i(d), their energy is sizeable 80\,meV below the average, we also find several configurations with $17$, $44$ and even $113\,$meV above it. This indicates that there might be specific local environments which are particularly favourable for the placement of an Mn atom and may thus lead to a kind of short-range order. These are, however, not straight-forward to identify from our calculations in the 28-atom cell, with its pseudo-disordered placement of Si, and thus a systematic comparison in larger cells is necessary, which is beyond the scope of the present work. However, we may conclude that one obvious parameter, the number of Si atoms in the closest neighbour shells of Mn, $n_{\mathrm{Si}}$, which is also given in Table \ref{tab:DFT-moments} does not correlate with the energy of the configuration.
	
	Substitutional Mn on the 96i sites enters a rather stable high moment state with an average magnitude of $2.6\,\mu_{\mathrm{B}}$, which is oriented oppositely to the ferromagnetically aligned Fe-moments. This is roughly half of a Bohr magneton larger than the average moment of Fe and corresponds to the smaller valence electron number e/a of Mn. However, the magnetic moments of Mn are considerably smaller compared to other magnetocaloric TM compounds such as Ni-Mn-based Heusler alloys~\cite{aolu2004}. Ferromagnetic configurations with Mn on 96i positions aligned parallel to Fe turned out to have significantly larger energy and were not considered further. The antiparallel orientation of Mn leads as a side effect to a slight reduction of the induced moments of La by 0.02\,$\mu_{\mathrm{B}}$ and Si by roughly 0.013\,$\mu_{\mathrm{B}}$, which is in agreement with the measurements presented in the preceding subsection. The Fe moments, in turn, reduce by 0.066\,$\mu_{\mathrm{B}}$ on average.
	
	If Mn is placed on the 8b sites, a high moment state can not be stabilized in the calculations. Instead, a low moment state with a magnitude below 1\,$\mu_{\mathrm{B}}$ is realized, which was previously also observed for Fe on the 8b sites \cite{Gruner2015,Gruner2017}. Our data cannot give a final conclusion about the orientation of the Mn-moment; thus antiparallel or non-collinear alignments may be possible as well.
	
	The low spin configuration of Mn on the Fe$_{\mathrm{I}}$ sites or the antiparallel orientation of Mn on the Fe$_{\mathrm{II}}$ sites results in a decrease in the average ground state volume of $0.25\,$\% and $0.33\,$\%, respectively, relative to the volume of LaFe$_{11.5}$Si$_{1.5}$, see Table\ \ref{tab:DFT-dist}. The volume change depends substantially on the position of the Mn on the inequivalent Fe-sites. Furthermore, the decrease in average volume does, however, not result in a homogeneous reduction of the nearest neighbour interatomic spacings, which are also listed in Table\ \ref{tab:DFT-dist}. 
	For the case that Mn is placed on the Fe$_{\mathrm{II}}$ (96i) sites, the average distance of La to all neighbours closer than $4\,$\AA{} reduces by $0.096\,$\% on average, for Fe$_{\mathrm{I}}$ to $0.19\,$\% and Fe$_{\mathrm{II}}$ to $0.11\,$\% all neighbours closer than $3\,$\AA{}. For Mn atoms on the 96i sites, which introduce the antiparallel moments, the average distance to all nearest neighbors reduces only by $0.067\,$\% compared to the nearest neighbor distances of Fe$_{\mathrm{II}}$ in LaFe$_{11.5}$Si$_{1.5}$.
	The origin of this reduced trend may be deduced to the significantly larger distances of Mn to the closest Si atoms, which expand by a considerable $2.5\,$\% compared to their Fe$_{\mathrm{II}}$ counterparts, which remain nearly constant in the presence of Mn despite the reduction of the total cell volume. This points out the paramount role of the Fe-Si bonding in stabilizing this alloy while Mn attempts to establish a larger distance to the surrounding Si. In fact, in Table\ \ref{tab:DFT-moments} those configurations of Mn on 96i sites reveal the lowest energy, which have no nearest Si neighbours or a particularly large separation. In turn, the distances between La and Si decrease by $0.27\,$\% on average. A comparison of the electronic density of states (not shown) does not exhibit signs of an increased hybridization between the partial contributions of the two elements. Therefore, we tentatively exclude stronger La-Si bonding as the origin. Instead, we rather ascribe this to the large spacing between La and its neighbours, which offers more space for adapting to structural modification arising from Mn-Si interactions as compared to Fe.
	
	When Mn is placed on the Fe$_{\mathrm{I}}$ 8b sites, the La-Si distances decrease to a similar extent, while the  Fe$_{\mathrm{II}}$-Si distances remain largely constant on average, while Mn-Si distances are again significantly larger. Here, the configuration with no nearest Mn-Si neighbours offers the lowest energy.

	\subsection{The microscopic geometric structure}
	
	Extended X-Ray absorption fine structure spectroscopy (EXAFS) measurements were performed at the Fe K, Mn K, and La L$_{3}$ edges for all Mn-concentrations at the beamline BM30B (ESRF). By performing EXAFS measurements of the different Mn-doped compounds, it is possible to follow the influence of the Mn concentration on the short-range ordering around the probed atom, especially unravelling the site occupation of the Mn atoms. Besides, it is possible to determine upcoming changes in the local environment of the probed element. The oscillatory fine structure following the element-specific absorption edge can be understood in terms of the EXAFS equation: 
	\begin{equation}
		\begin{split}
			\chi(k) = &\sum_{j}^{\,} S_0^2(k) N_j f_j(k) 
			\exp\left(-2 \frac{R_j}{\lambda(k)}\right) \exp\left(-2\sigma^2_jk^2\right)\\
			&\cdot \frac{\sin\left(2kR_j - \dfrac{4}{3}C_{3,j}(T)k^3+ \delta_j(k)\right)}{2kR^2_j}
		\end{split}
		\label{eq: EXAFS}
	\end{equation}
	Hereby, the oscillatory part is given through the last term, where $k$ is the wave number of the photoelectron, $R_j$ is the distance to neighbour atoms in shell $j$ and $\delta_j(k)$ originates from the finite size of the scattering potential giving rise to a phase shift of the scattered wave. The oscillation strength mainly depends on the number of identical atoms per shell $N_j$. Depending on the scattering element, the shape of the scattering potential changes in an energy-dependent manner, considered through the backscattering amplitude $f_j(k)$. The finite lifetime of the photoelectron leads to a damping of the fine structure at larger wavenumbers $k$. It can be described by an energy-dependent mean free path $\lambda(k)$ of the photoelectron. In addition, structural disorder leads to additional damping of the fine structure in the form of the Debye-Waller factor and is characterised through the mean square relative displacement $\sigma^2_j$ which can be further divided in the sum of the static contribution $\sigma^2_\text{static}$ and the temperature-dependent dynamic contribution $\sigma^2_\text{dynamic}(T)$ arising from lattice vibrations. In order to ensure minimal contributions of dynamic disorder, the measurements were performed at $T=20$\,K, so we assume to be in the regime of zero-point vibrations that can be well described within a Debye model. The third cumulant $C_3(T)$ reflects anharmonic lattice vibrations~\cite{Fornasini2001} that occur at elevated temperatures. Due to the low measurement temperature, anharmonic contributions can be neglected within the scope of this study~\cite{Sanson2009}. As equation (\ref{eq: EXAFS}) indicates, EXAFS is a technique averaging all absorber atoms. Thus, one spectrum arises from the sum of the probed element's lattice sites and the respective local environment. 
	
	\begin{figure*}[h!]
		\centering
		\includegraphics[width=\linewidth]{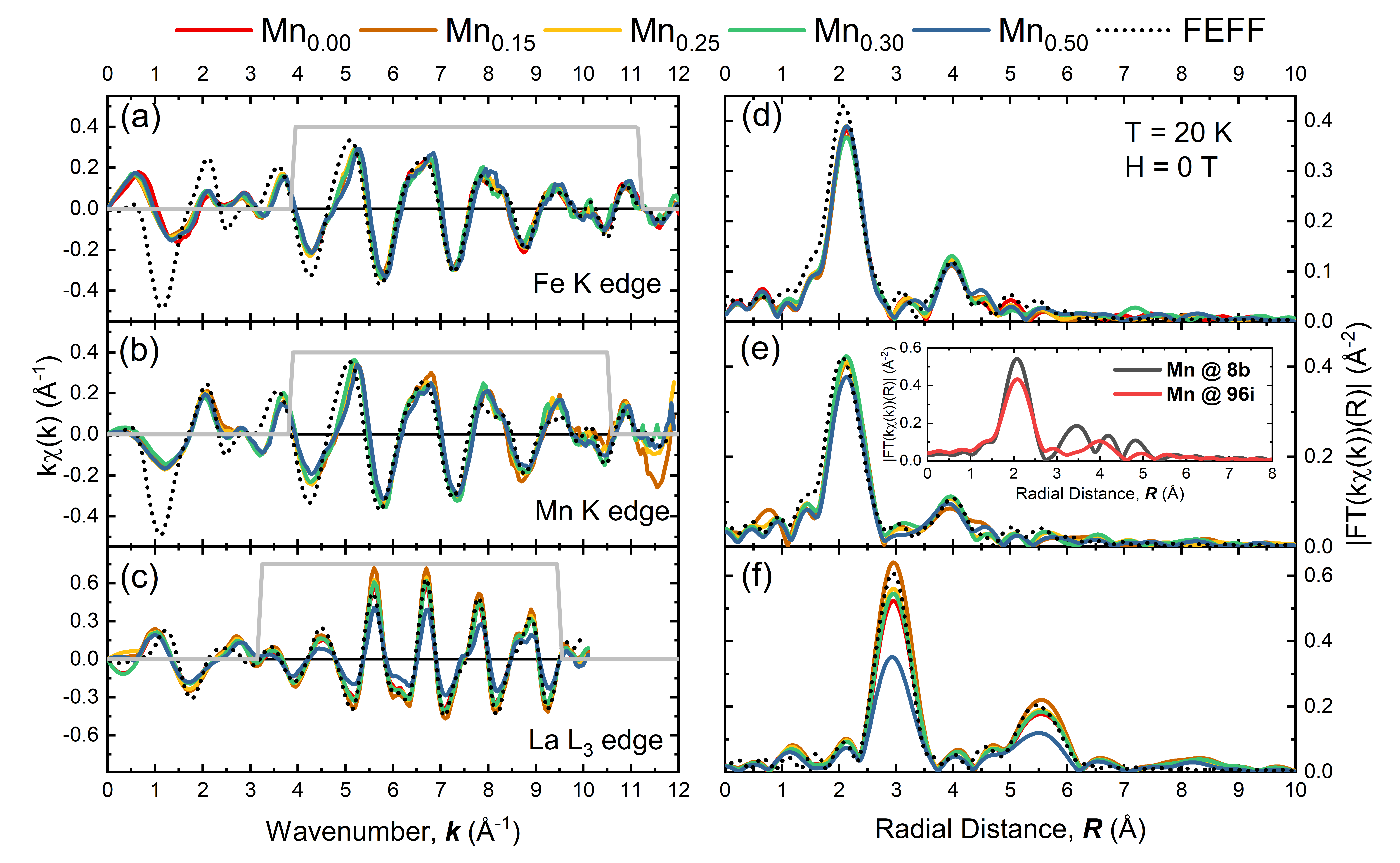}
		\caption{EXAFS oscillations $k\chi\left(k\right)$ for LaFe$_{\text{11.6-x}}$Mn$_{\text{x}}$Si$_{\text{1.4}}$. at the Fe K edge (a), Mn K edge (b), and La L$_3$ edge (c) measured at 20\,K in the absence of a magnetic field. (d)-(f) show the corresponding Fourier transformations $\left|FT\left(k\chi\left(k\right)\right)\right|\left(R\right)$ that was performed with the indicated window (grey line in subfigures (a)-(e)). The black lines depict ab-initio FEFF-calculations with an assumed Debye temperature $\theta_{\text{D}}$ of 380\,K and a measurement temperature of 20\,K for the lowest Mn-concentration. The inset in subfigure (e) depicts the Fourier transform from FEFF calculations, where Mn occupies the 8b site (black line) or 96i site (red line).}
		\label{fig:EXAFS-combined}
	\end{figure*}
	
	Results for Mn-concentrations Mn$_{0.00}$, Mn$_{0.15}$, Mn$_{0.25}$, Mn$_{0.30}$ and Mn$_{0.50}$ of the measurements are shown in Figure~\ref{fig:EXAFS-combined}. The fine structure oscillations at the Mn and Fe K edge indicate that Mn occupies both the 96i and 8b sites. Furthermore, FEFF calculations describe the experimentally obtained EXAFS spectra at the La L$_3$ and Fe K for the initial sample and the Mn K edge with a slightly adapted unit cell to model the Mn$_{0.50}$ doped sample. The FEFF calculations are depicted as black dotted lines, while the measurements are coloured, depending on the Mn concentration. For the FEFF calculations, a correlated Debye model~\cite{Rehr2010} with a Debye-Temperature $\Theta_\text{D}$ of 380\.K~\cite{Landers2018} was used to take dynamical disorder effects into account, arising from finite temperatures. Additionally, a static disorder of $\sigma^2_\text{static} = 0.001\,$Å was taken into account, representing slight lattice disorder in an expected range. Subfigures~\ref{fig:EXAFS-combined} (a) - (c) show the oscillatory fine structure of all samples at the mentioned three edges. With these measurements and calculations, we aim to resolve the occupation sites of the Mn atoms in the NaZn$_{13}$ structure and then continue to discuss the influence of Mn doping on the short-range order. 
	
	The experimental fine structures of the Mn K edges and Fe K edges exhibit similarities in frequency and fine structure. Site-resolved calculations of the 8b and 96i sites for the Fe K edge and Mn K edge indicate that Mn atoms cannot be solely found on the 8b sites. The site-resolved Fourier transforms of 8b and 96i sites (see inset of Figure~\ref{fig:EXAFS-combined}(e)) reveal the first backscattering shell at R=2.1\,\AA{} corresponding to a bond length of approximately 2.4\,\AA{}. Assuming that Mn occupies the 8b site, the Fourier transform of the corresponding FEFF calculation depicts a clear contribution at 3.5\,\AA{} that is not present in our experimental results or FEFF calculations for Mn occupying the 96i site. The missing contribution at R=3.5\,\AA{} indicates that the majority of the Fe and Mn fine structure originates from 96i-surroundings. Due to the small spectral contribution of the 8b sites, it is not possible to state whether the Mn atoms occupy both sites, the 8b and 96i sites or only the 96i sites. The derived occupation is consistent with previously published Mossbauer results~\cite{Makarov2015} and our first-principles calculation (discussed above). In contrast, a recent investigation of Cr-doped La(Fe,Si)$_{\text{13}}$~\cite{MorenoRamrez2019} reveals that Cr only occupies the 8b site. At the same time, the occupation of the 96i-site is energetically unfavoured by 0.31\,eV per Cr atom compared to the 8b-occupation~\cite{MorenoRamrez2019}. This indicates a clear difference of the site occupation between Mn and Cr.
	
	By comparing the different Mn-concentrations for each edge, it can be seen that while for the Fe K edge (Subfigure~\ref{fig:EXAFS-combined} (a)) and the Mn K edge (Subfigure~\ref{fig:EXAFS-combined} (b)), there is no change in the oscillatory fine structure visible beyond the error of measurement, the La L$_3$ edge shows a clear decreasing envelope of its fine structure with increasing Mn-concentration. The oscillatory frequency remains the same, which denotes no change in the distance $R_j$. The decrease in the envelope indicates increasing structural disorder at a constant temperature, an effect of rising static disorder. This effect is again visible by looking at the Fourier transform of the EXAFS signal for the respective edges (see subfigures~\ref{fig:EXAFS-combined} (d) - (f)). The Fourier transform of the EXAFS signal is connected to the pair distribution function, and therefore, each contribution represents atoms in a neighbouring shell. While the Fourier transforms of the Fe K and Mn K edge lay on top of each other (within experimental accuracy), the La  L$_3$ edge shows an intensity decrease with increasing Mn concentration. \\
	
	\begin{figure}[h!]
		\centering
		\includegraphics[width=.75\linewidth]{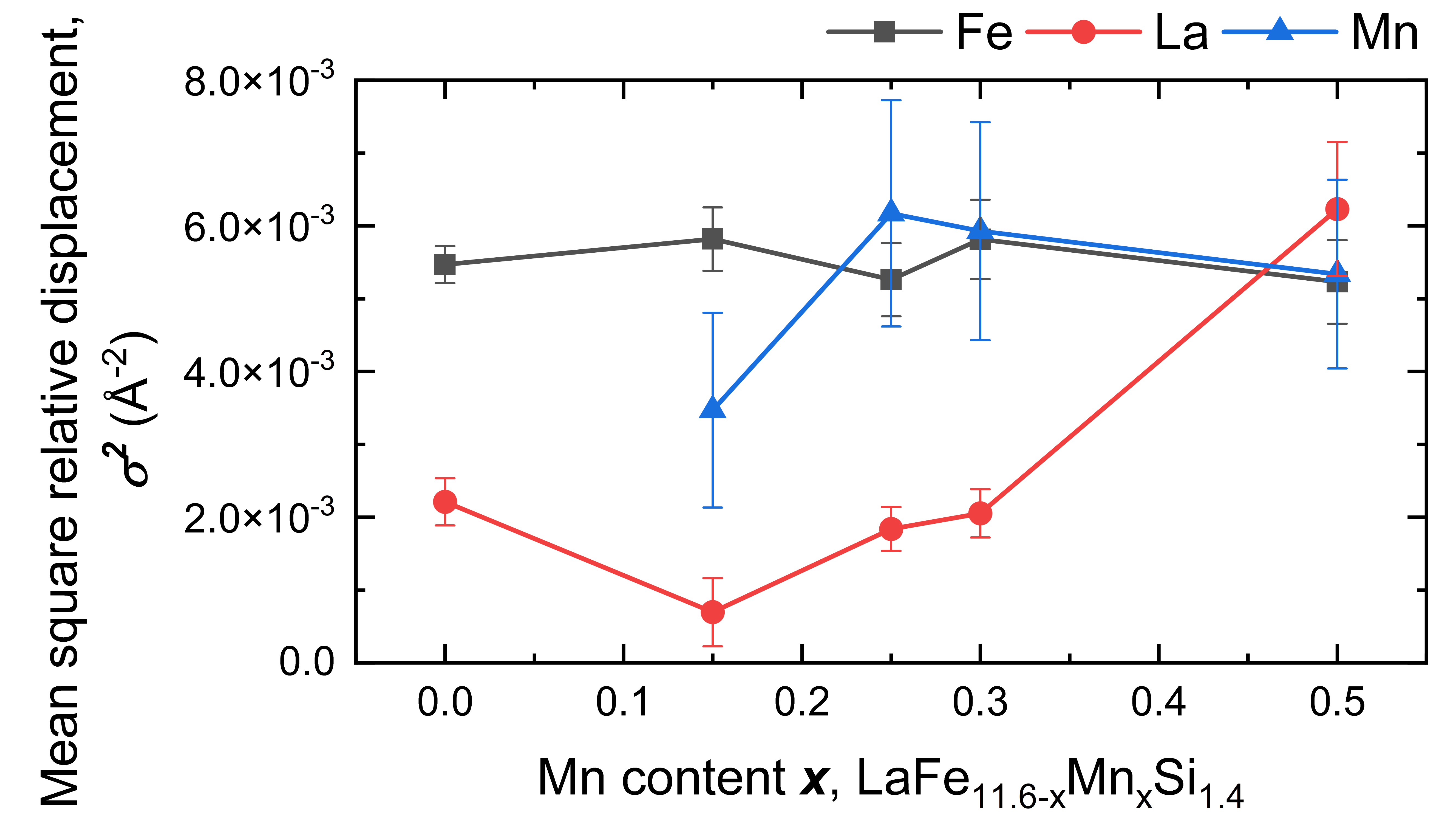}
		\caption{Extracted mean square relative displacements $\sigma^2$ for the first backscattering shell of La, Fe, and Mn. Fits performed for measurements taken at T = 20\,K.}
		\label{fig:EXAFS-DW}
	\end{figure}
	
	To further analyze the effect of structural disorder, a modelling of the first backscattering shell has been employed with single scattering paths within the ARTEMIS and Larch package~\cite{Ravel2005,Newville2013}. For this model, we assume a uniform mean square relative displacement (MSRD) $\sigma^2$. Based on these calculations, we can extract variations of the structural disorder for the different Mn concentrations (see Fig.~\ref{fig:EXAFS-DW}). The overall mean square relative displacements (static and dynamic disorder) of the Fe and Mn K edges are constant within the error for all Mn concentrations. However, the La L$_3$ edge presents a strong increase of the MSRD with increasing Mn concentration. These results depict that the Fe and Mn environment presents no observable response to the increasing Mn concentration, representing no drastic change in the local structure or increasing disorder in the local environment around the Fe and Mn absorbers, respectively.
	
	In contrast, the drastic increase of the mean square relative displacement seen for La L$_3$ edge needs to be discussed further. We assume that phonons and zero-point vibrations of Fe and Mn do not differ substantially due to their similar atomic weight. Therefore, a La-Fe or a La-Mn scattering path's mean square relative displacement should be identical and not depend on the Mn concentration. On the other hand, a La-Si path possesses a different MSRD due to the small weight of Si compared to La, Fe, or Mn and would show larger variations. As discussed in section~\ref{ch:DFT}, Mn-Si bonds possess increased interatomic distances compared to other Fe$_{\mathrm{II}}$-Si ones (see Table~\ref{tab:DFT-dist}). Besides, our DFT calculations predict a distribution of Mn-Si and La-Si distances, leading to the increased mean square relative displacement $\sigma^2$ of Mn and La. However, due to the high amount of degenerated single scattering paths in the first backscattering shell of La (22 paths) and similar backscattering amplitudes for La-Si, La-Fe and La-Mn, it is impossible to experimentally distinguish these spectral contributions further. Since EXAFS averages all local environments of the probed element, the increased structural disorder is only reflected at the La edge and not at the TM K edges (e.g. through a TM-Si scattering path) since Fe or Mn, in principle, occupy 93 (85 Fe II and 8 Fe I) out of 112 positions per unit cell. In contrast, Si only occupies 11 of these, with La having eight individual positions. Due to the investigated compound's small Si content, each Fe atom possesses only one Si as a nearest-neighbour on average (see Table~\ref{tab:DFT-moments}). In contrast, all La-atoms have a higher probability for Si atoms in their closest surroundings. Therefore the La surrounding is more sensitive to the structural disorder caused by the Mn-Si bonds.

	\section{Summary}
	In summary, we combined element-specific XMCD, EXAFS, and first-principles calculations to follow the evolution with element-specific contrast of the magnetic moments and the lattice structure in La(Fe,Si)$_{\text{13}}$ with increasing Mn concentration. By employing XMCD at the Fe K and La L$_{2,3}$ edges, we could reveal that the 3d Fe moment reduces -- leading to a decreased magnetisation. Furthermore, DFT calculations indicate that Mn predominately occupies the 96i site with an average moment of 2.6$\,\mu_{\mathrm{B}}$ that couples antiferromagnetically to the otherwise ferromagnetically aligned Fe moments~\cite{Gercsi2015,Makarov2015} and is an additional source for a decreased magnetisation that we observe in conventional magnetometry. Besides, DFT calculations in our 28-atom cell indicate the presence of specific local environments that are favourable for Mn and might lead to a kind of short-range order, which does not only depend on the amount of Si nearest-neighbours. 
	
	Looking at the structural response, we can see that the introduction of Mn on 96i-sites reduces the volume and, in addition, leads to an inhomogeneous reduction of nearest-neighbour interatomic spacings. Here, we want to highlight the occurrence of relatively large Mn-Si distances that are 2.5\,\% larger compared to other Fe$_{\mathrm{II}}$-Si bonds (see Table~\ref{tab:DFT-dist}). These large Mn-Si bond lengths lead to a variation of La-Si spacings, resulting in structural disorder that we observe in the La environment through EXAFS but not in the Fe and Mn surroundings. The drastic increase of structural disorder with increasing Mn content might be a reason for the formation of additional secondary phases besides the  NaZn$_{13}$ structure observed in previous experimental studies~\cite{Krautz2014}.

	\section{Acknowledgements}
	This~work was funded by the Deutsche Forschungsgemeinschaft (DFG, German Research Foundation) within TRR~270 (subprojects~A03, B01, B05, and~B06), Project-ID~405553726. We thank the ESRF for allocating beamtime at ID12 and BM30 and the financial support through projects HC-3255 and MA-4020. Calculations were performed at the MagnitUDE high performance computing system of the Center of Computational Sciences and Simulation (CCSS) at the University of Duisburg-Essen (DFG INST 20876/209-1 and 20876/243-1~FUGG).
	
	\bibliography{main}
\providecommand*{\mcitethebibliography}{\thebibliography}
\csname @ifundefined\endcsname{endmcitethebibliography}
{\let\endmcitethebibliography\endthebibliography}{}

\end{document}